\documentclass[12pt]{article}
\usepackage{epsfig, graphics, subfigure}
\usepackage{amsfonts}
\usepackage{eurosym}
\usepackage{mathtools, bm}
\usepackage{amssymb, bm}
\usepackage[latin1]{inputenc}

\def\bkR{{\rm I\kern-.17em R}}
\def\QED{\begin{flushright} QED \end{flushright}}

\def \1n{1\hskip -3pt \mbox{N}}
\def \eR{I\hskip -3pt R}
\def \Frac {\displaystyle \frac }
\def \Int {\displaystyle \int }
\def \Sum {\displaystyle \sum }

\begin{document}

\begin{titlepage}
\thispagestyle{empty}

\title{Structural Modelling of Dynamic Networks and Identifying Maximum Likelihood}
\author{GOURIEROUX, C., $^{(1)}$  and J., JASIAK $^{(2)}$}
\date{\today}

\maketitle \vspace{10cm}

\noindent We thank X. D'Haultfeuille, M. Henry and A. Monfort for helpful comments.

\addtocounter{footnote}{1} \footnotetext{University of Toronto, Toulouse School of Economics and CREST.}
\addtocounter{footnote}{1} \footnotetext{York University.}

\end{titlepage}

\begin{center}
\textbf{Structural Modelling of Dynamic Networks and Identifying Maximum Likelihood} \vspace{1em}

 Abstract
 \end{center}\vspace{1em}

This paper considers nonlinear dynamic models where the main parameter of interest is a nonnegative matrix
characterizing the network (contagion) effects. This network matrix is usually constrained either by assuming a limited number of nonzero elements (sparsity), or by considering a reduced rank approach for nonnegative matrix factorization (NMF). We follow the latter approach and develop a new probabilistic NMF method. We introduce a new Identifying Maximum Likelihood (IML) method for consistent estimation of  the identified set of admissible NMF's and derive its asymptotic distribution. Moreover, we propose a maximum likelihood estimator of the parameter matrix for a given non-negative rank, derive its asymptotic distribution and the associated efficiency bound.

\vspace{1em}

 \textbf{Keywords~:} Network, Nonnegative Matrix Factorization, Set Identification, Heterogeneity, Identifying Maximum Likelihood (IML), Alternating Maximum Likelihood (AML).
 \newpage

 \section{Introduction}

 The network effects are commonly represented by a non-negative matrix $A$ of dimension $(n,m)$. There exists a large literature on network models. It can be divided into research strands that differ with respect to the objective of analysis, which is either prediction-oriented or structural, as well as with respect to the assumptions imposed on the non-negative matrix $A$. These assumptions may concern the 0 and 1 entries of adjacency matrices, or the positive elements of incidence matrices and their "reduced rank".
In applications to either contagion, transmission, social interaction, or spillover effects, the elements of each row of matrix $A$ can be constrained to  sum up to one. Then, each row of matrix $A$ is interpreted as a conditional probability distribution and that matrix is called a transition, or migration matrix.

 The prediction-oriented techniques are applied to image analysis, facial recognition and machine learning. The dimensions $n$ and $m$ of matrix $A$ are very large, and the primary goal of the method is to reduce its complexity. Hence, adjacency matrices with a small number of links, i.e. entries equal to 1 are used to obtain sparse matrices  $A$. Also, incidence matrices are chosen so that non-negative matrix factorization (NMF) with a rather small reduced rank can be applied [Berman, Plemmons (1994)].

 The methods applied to network analysis in econometrics are more structural and parametric [see Manski (1993) for the reflection problem, Blume et al. (2011), de Paula (2017) for a survey and the references therein]. 
The models often represent an equilibrium, such as:

$$Y_t = A Y_t + D X_t + u_t, $$

\noindent where $Y_t = (y_{1,t}, ...., y_{n,t})$ are the individual observations at time $t$, and $X_t$ are the explanatory variables. At the equilibrium, a simultaneity arises because the variable $Y$ appears on both sides of the equation. The dimensions $n,m$ of matrix $A$ are much smaller than in the image analysis. Moreover,  in order to obtain a simple parametric model, matrix $A$ is often defined as a parametric nonnegative combination $A = \Sum^L_{l=1} \alpha_l A_l, \; \alpha_l \geq 0,\; l=1,\ldots, L$, of known  network matrices, which is a crucial assumption in this literature [see e.g. Bramoulle et al. (2009), Lee et al. (2010), De Giorgi et al. (2010), Cohen-Cole et al. (2014), Blume et al. (2015)].

 The aim of this paper is to fill the gap between the prediction-oriented and the structural methods by introducing the identifying Maximum Likelihood method, a Maximum Likelihood (ML) estimator of the set of admissible NMFs as well as an ML estimator of matrix $A$, in an extended class of dynamic probabilistic models.
 
We introduce the class of dynamic parametric models with interaction matrices of reduced rank, and discuss various examples, such as the static models used in the machine learning literature, the dynamic panel models for individual qualitative histories, and the multivariate dynamic Poisson model with contagion used in epidemiology, as special cases of that extended class of models. The NMF of matrix $A$ in this extended class of models may involve not only
the directional factors, but also a latent heterogeneity distribution. In general, there exist multiple admissible factorizations of a given true network matrix $A_0$ with or without zero entries. We describe analytically the identified set for nonnegative ranks $K=1$, and $K=2$ of matrix $A$ and show how it can be parametrized in the general case $K \geq 2$. The proposed statistical inference methods rely on the Identifying Maximum Likelihood approach which is introduced to estimate the set of NMF's and to derive the asymptotic distribution of the estimated identified set. In this context, the estimation of a NMF with the most concentrated latent heterogeneity is also considered. Moreover, we propose a maximum likelihood estimator of the nonnegative matrix $A$, given its nonnegative rank, and we derive the efficiency bound for the rank constrained matrix parameter. 
 
Our approach differs from the methods used in the existing network literature with respect to the following 1) In our paper, the network is examined in a nonlinear dynamic framework, allowing us to distinguish the short and long run effects. Our approach is applicable to static linear models considered in the existing literature, which arise as special cases;  2) The matrix $A$ is compatible with the existing empirical applications that mainly concern the semi-aggregated level networks with
matrices $A$ without zero elements (intricate network) and the diagonal (resp. off-diagonal) elements representing the within (resp. between) segment connections; 3) In our paper, matrix $A$ is assumed to satisfy a factor decomposition into non-negative factorial directions. To solve the partial identification problem, we derive analytically the identified set and show that it can be locally parametrized. In particular, the identified set is not defined from inequality moment restrictions, as it is commonly assumed in micro-econometric partial identification literature. Instead, the Identifying Maximum Likelihood (IML) method is introduced in this paper, allowing for sharp estimation of the identified set and of the rank constrained matrix $A$.

The paper is organized as follows.
 In Section 2, we introduce the class of dynamic parametric models with interaction matrices of reduced rank. Section 3 discusses the identification of a NMF of matrix $A$. Statistical inference is developed in Section 4.  Section 5 presents the link with the nonparametric approaches to partial identification. Section 6 concludes. 

 \section{Parametric Model with Interaction Matrix}
 \setcounter{equation}{0}\def\theequation{2.\arabic{equation}}

 \subsection{The model}

We consider a set of observations $Y_t, t=1,\ldots, T$, that can be scalars, vectors, or matrices. We assume that $(Y_t)$ is a stationary Markov process and introduce a parametric model for the conditional distribution of $Y_t$ given its lagged values, with a conditional probability density function (p.d.f.)~:

 \begin{equation}
   l(y_t | \underline{y_{t-1}}) = l(y_t|y_{t-1} ; A),
 \end{equation}

 \noindent where $A$ is an unknown non-negative matrix $A \geq 0$ of dimension $(n,m)$, which has non-negative entries.

 We make the following assumption~:\vspace{1em}

 \textbf{Assumption A.1~:}

 i) The parametric model is well-specified, with a true value $A_0$ of matrix parameter $A$.

 ii) The process $(Y_t)$ is strictly stationary, geometrically ergodic.\vspace{1em}

 It is well-known that a nonnegative matrix $A$ can be factorized as~:

 \begin{equation}
   A = B C',
 \end{equation}

 \noindent where $B$ (resp.$C$) have dimensions $(n,K)$ [resp. $(m,K)$] and are nonnegative~: $B \geq 0, C \geq 0$. Among the multiple non-negative factorizations available, some correspond to a minimal order $K$, called the nonnegative rank of matrix $A$ and denoted by $Rk_+ (A)$. The nonnegative rank of $A$ is always larger or equal to the rank of $A$.

 A nonnegative matrix factorization (NMF) can be written under different equivalent forms. Let  $\beta_k, k=1,\ldots, K$ (resp. $\gamma_k, k=1,\ldots,K$) denote the columns of $B$ (resp. $C$). We have~:

 \begin{equation}
   B = (\beta_1, \ldots, \beta_K), C = (\gamma_1,\ldots, \gamma_K),
 \end{equation}

 \noindent and then~:

 \begin{equation}
   A = \Sum^K_{k=1} \beta_k \gamma'_k\; \mbox{with}\; \beta_k \geq0, \gamma_k \geq 0, \forall k.
 \end{equation}

\noindent This provides a decomposition of $A$ as the sum of $K$ non-negative matrices of rank 1.\vspace{1em}

In structural models, we can be interested not only in the true value $A_0$, but also in a true NMF~: $B_0 C'_0 = \Sum^{K_0}_{k=1} \beta_{0,k} \gamma'_{0,k}$, that generates $A_0$. We introduce the additional assumptions~:\vspace{1em}

 \textbf{Assumption A.2~:}

 i) The nonnegative rank $K_0=Rk_+ (A_0)$ is known.

  ii) The true matrix $A_0$ is asymptotically identifiable, that is~: the optimization problem $\max_{A} E_0 \log l(Y_t | Y_{t-1}; A)$ has the unique solution $A=A_0$.

 iii) The vectors $\beta_{0,k}, \gamma_{0,k}, k=1, \ldots, K,$ have strictly positive entries.\vspace{1em}

 \textbf{Remark 1~:} The analysis of this paper is easily extended to parametric conditional models including also observed exogenous variables $X_t$, or additional parameters $\theta$, that is to models of the type~:

 $$
 l (y_t | \underline{y_{t-1}}, \underline{x_t}) = l (y_t | y_{t-1}, x_t ; A_0, \theta_0),
 $$

\noindent that is to dynamic panel models with covariates.

 \subsection{Examples}

In general the NMF is applied without assuming a probabilistic structure and our objective is to extend to NMF, what has been done for principal component analysis by Tipping, Bishop (1999). The examples below show models that could be introduced for different types of applications as facial recognition, epidemiology, or credit risk. These models have to account for the nonnegativity of the observations $Y_t$, usually encountered in practice.

 \subsubsection{Static model}

 The static models assume that observations $Y_t, t=1,..,T$ are independent and identically distributed   (i.i.d.). These models are commonly used for image analysis [see Lee, Seung (1999) for the first application to learning the parts of objects] under a non-probabilistic approach. In this application, the observations are matrices $Y_t = (Y_{i,j,t}),$ where $(i,j)$ denote the coordinates of a point in picture $t$ and the value $Y_{i,j,t}$ provides the pixel intensities associated to coordinates $i,j$ and picture $t$. The pixel intensity can be measured on either a discrete, or continuous scale. Then, the  considered model:

 \begin{equation}
   l(y_t | y_{t-1};A) = \Pi^n_{i=1} \Pi^m_{j=1} f(y_{i,j,t}; a_{i,j}),
 \end{equation}

 \noindent is based on a family of probability density functions (p.d.f.) $f(y; a)$ with nonnegative argument $y$ and nonnegative scalar parameter $a$. For the Poisson and exponential p.d.f., the specification in (2.5) simplifies and leads to a generalized linear model (GLIM) [see McCullagh, Nelder (1989) for GLIM and Collins et al. (2002) for its use for factor analysis].

In other applications, network matrices can be observed and analyzed by static models. For instance, we can consider a set of individuals $i=1,\ldots, L$ and observe at time $t$ the number $y_{i,j,t}$ of messages sent by individual $i$ to individual $j$. The i.i.d observations $y_t$ can also be matrices containing the investments of bank $i$ in industrial sector $j$ at time $t$ , the gravity matrices summarizing the international trades between countries [Chen et al. (2021), Section 6], or the matrices representing the numbers of stocks of firms head-quartered in city $j,j=1, \ldots, m$ and selected by mutual fund manager $i, i=1,\ldots,n$ at time $t$ [Hong, Xu (2014)].

 \subsubsection{Panel of individual qualitative histories.}

 Let us consider a panel of $L$ individuals. Each individual is characterized by a qualitative state $i=1,\ldots, n,$ that can be observed at any time $t$. The qualitative individuals histories can be quantified and represented by $n$-dimensional vectors $(Y_{l,t}, t=1,\ldots,T), \; l=1,\ldots, L$, where $Y_{l,t}$ has entries  that sum up to 1.

\noindent The dynamic model of  can be defined by assuming that~:

 i) the individual histories are independent;

 ii) each individual history corresponds to a Markov chain with transition matrix $A$.\vspace{1em}

\noindent  The above assumptions imply that the population of interest is homogeneous. Under these assumptions the individual histories can be aggregated without a loss of information and replaced by the counts of individuals in each state~:

 \begin{equation}
   Y_t = \Sum^L_{l=1} Y_{l,t}.
 \end{equation}

\noindent Then the sequence of multivariate counts $Y_t$ is also a Markov process with the conditional p.d.f. obtained by considering the convoluate of different multinomial distributions~:

 \begin{equation}
   l(y_t |y_{t-1}; A) = \Pi^n_{i=1} l_i (y_{i,t} | y_{i,t-1}; a_i),
 \end{equation}

 \noindent where $l_i$ denotes the p.d.f. of the multinomial distribution $M (y_{i,t-1};a_i)$ and $a_i$ denotes the $i^{th}$-row of the transition matrix $A$.
 
 In this example, we have replaced the individuals by homogeneous segments. For example, the corporates can be replaced by industrial sectors, or households by different classes of ages. Then, the diagonal elements of matrix $A$  depict the interactions within a segment and the off-diagonal elements of $A$ represent the interactions between the segments. In general, the diagonal elements $a_{ii}$ are not equal to zero.

 \subsubsection{Dynamic model for a non-negative random vector}

 Let us consider a parametric model for a non-negative random vector $l(y;\theta)$, where $y$ and the parameter vector $\theta$ have equal dimension $n$, and both vectors $y$ and $\theta$ are nonnegative. Then, a dynamic model for $(y_t)$ can be defined as~:

 \begin{equation}
   l (y_t | y_{t-1} ; A) = l(y_t;A y_{t-1}),
 \end{equation}

 \noindent where the contagion matrix $A$ is of dimension $(n,n)$ and nonnegative.

 When the parametric model represents the dynamics of $n$ independent Poisson variables, we get a multivariate Poisson autoregressive model with~:

 \begin{eqnarray}
   l(y_t | y_{t-1} ; A) & = & \Pi^n_{i=1} \left[ \Frac{1}{y_{it !}} \exp (-a_i y_{t-1}) (a_i y_{t-1})^{y_{it}}\right] \nonumber \\
   &=&\Pi^n_{i=1} \left( \frac{1}{y_{it !}}  (a_i y_{t-1})^{y_{it}} \right) \exp (-e' A y_{t-1}) ,
 \end{eqnarray}

 \noindent where $a_i$ is the $i^{th}$ row of matrix $A$ and $e$ is the vector with unitary elements.

 \noindent This specification differs from an exponential specification:

 $$
 y_{it} | y_{t-1} \sim \mathcal{P} [\exp (a_i y_{t-1})],
 $$

 \noindent considered in Chen et al. (2021), Example 3, and Section 6. This alternative specification does not require matrix $A$ to be nonnegative. However, while the dynamic of model (2.9) is compatible with the stationarity of the process $(y_t)$, the above alternative approach of Chen (2021) leads to explosive trajectories due to the exponential transform.

 When the parametric model represents the dynamic of $n$ independent exponential variables, we get a multivariate exponential autoregressive model:

 \begin{equation}
   l(y_t | y_{t-1} ; A) = \Pi^n_{i=1} \left[ (a_i y_{t-1}) \exp (-a_i y_{t-1} y_{it})\right].
 \end{equation}
 \vspace{1em}

 \noindent This dynamic model can be used to study the joint evolutions of the gross domestic product in a set of $L$ countries. Then, matrix $A$ is unobserved and needs be estimated under some mild constraints to provide a proxy of an international trading network.\vspace{1em}

 \textbf{Remark 2~:} The dynamic specification (2.8) can be extended to~:

 $$
 l(y_t|y_{t-1}; A) = l(y_t ; A z_{t-1}),
 $$

 \noindent where $z_{t-1}$ is a nonnegative vector function of $y_{t-1}$. Such a transformation appears in  structural models used in epidemiology and in other applications such as the analysis of cyberattacks [Fahrenwaldt et al. (2018), Hillairet, Lopez (2020)], adoption of a new technologies [Brock, Durlauf (2010), Blume et al. (2011)] and the Susceptible-Infected-Recovery (SIR) model with multiple transmissions where the components of $y$ are the counts of infected individuals [resp. of the new adoptions of the product] in different segments of the populations. In the SIR model,  $z$ is a quadratic function of $y$ [see e.g. Gourieroux, Jasiak (2022)].

\subsubsection{Contagion of defaults}

The structural models of corporate defaults assume that a default arises when the liability of a firm falls below its asset value. When only the defaults are observed, the individual assets and liabilities are latent nonnegative variables which can be represented by a dynamic network model (2.8):

\begin{equation}
  \tilde{l} (z_t |z_{t-1}; A) = \tilde{l} (z_t ; A z_{t-1}),
\end{equation}

\noindent where $z_t$ has dimension $2L$ and $L$ is the number of firms. The elements of $z_t$ represent the assets and liabilities of firms. For example, $z_{1lt}, z_{2lt}$ are the characteristics of firm $l$ at date $t$. The observed variables are defaults, represented by vector $y_t$ of dimension $L$ and defined by~:

\begin{equation}
  y_{lt} = 1, \; \mbox{if}\; z_{1lt} > z_{2lt}, = 0, \; \mbox{otherwise}.
\end{equation}

\noindent Equations (2.11) and (2.12) can be interpreted as a state space model. From the state equation (2.11) and the measurement equation (2.12) it follows that the transition of the observed variables is $l (y_t | \underline{y_{t-1}} ; A)$, involving the complete histories of default of all firms.

\medskip

All dynamic models given above have nonlinear dynamics due to the constraints on the values of observed variables. They are also nonlinear with respect to parameter $A$. These nonlinear features distinguish the class of models considered in this paper from the major part of literature on networks in econometrics. \footnote{"Identification and measurement of network phenomena has drawn attention in fields as diverse as macroeconomics, finance not discussed in the review" [De Paula (2017)]. These other fields with nonlinear dynamic models are also left out of the recent survey on partial identification by Kline, Tamer (2022). The nonlinear dynamic models in our paper are introduced for application to these other fields that  also include ecological economy, monetary economy and epidemiology.}. 

The identification in this extended class of models eliminates the use of powers of matrix $A$. Indeed, matrices $A, A^2, A^3$ no longer not play a special role, as the conditional distribution of $y_t$ given $y_{t-h}$ does not necessarily depend on $A$ through $A^h$ only [see e.g. de Paula (2017) p 272 for a discussion]. In fact all information on $A$ is captured by the lag 1 of variables in the likelihood function. 

Moreover, the response of a nonlinear dynamic system to a shock at date $t$ depends on the current environment of $y_t$. The size of the shock effect depends on the environment, so that even a small shock can have a large impact. This has important consequences concerning the treatment of small values of the elements $a_{ij}$ of matrix $A$. More precisely, even if $a_{ij}$ is small, this element should not to be set equal to zero artificially, by applying, for example, an automatic LASSO penalty [see e.g de Paula et al. (2020) p285-286]. Especially when the dynamic system is close to a tipping point, a small connection can become the source of a significant change in the system, as observed in the histories of corporate defaults, chains of business failures, or inter-bank liquidity shortages.

\subsection{Latent Heterogeneity and Ranking}

\subsubsection{Alternative parametrization}

The NMF (2.4) can be normalized and written alternatively as~:

\begin{equation}
  A = a \Sum^K_{k=1} \pi_k \beta^
  *_k \gamma^{*'}_k,
\end{equation}

\noindent where $a= e' A e = \Sigma_i \Sigma_j a_{i,j} >0, \pi_k \geq 0, k=1,\ldots, K$, with $\Sum^K_{k=1} \pi_k = 1, \beta^*_k \geq 0, \gamma^*_k \geq 0, k=1, \ldots, K$ with ~: $\beta^{*'}_k e = \gamma^{*'}_k e = 1, k=1,\ldots, K$.\vspace{1em}

In this decomposition $\pi = (\pi_1, \ldots, \pi_K)', \beta^*_k, \gamma^*_k, k=1, \ldots, K$ can be interpreted as discrete probability distributions. More precisely, the normalised matrix $A/a$ can be interpreted as a joint probability distribution, and its decomposition $\Sum^K_{k=1} \pi_k \beta^*_k \gamma^{*'}_k$ as a mixture of independent joint distributions. In this respect, our analysis is linked to the literature on partial identification of finite mixtures [see e.g. Henry et al. (2014) and Section 5]. When $K=1,$ the NMF becomes~: $A = a \beta^*_1 \gamma^{*'}_1.$

The representation (2.13) solves the identification issue due to the identification of factorial directions up to positive scalars. It also allows to bound the set of NMF's written under this form.

\subsubsection{Rankings}

To motivate the alternative parametrization (2.13), let us consider a dynamic model of count variables for epidemiology. The dynamic contagion model (2.8) can be considered, where the components of $y_t$ are the count of infected individuals in $L$ homogenous segments of the population. In addition, let us assume that \footnote{This constraint on the conditional mean implies $y_t = Ay_{t-1} + u_t$, where $u_t$ is a martingale difference sequence with $E_{t-1} (u_t) = 0$. It does not imply a linear dynamic model~: $y_t = A y_{t-1} + u_t$, where $u_t$ is a strong (i.i.d.) white noise. Hence, it is inadequate for the analysis of nonlinear shock effects, i.e. nonlinear impulse response functions.}~:

\begin{equation}
  E_{t-1} y_t = A y_{t-1}.
\end{equation}

\noindent If $K=1$, we get~: $E_{t-1} y_t = a \beta^*_1 \gamma^{*'}_1 y_{t-1}, $ or equivalently~:

\begin{equation}
  E_{t-1} y_{i,t} = \Sum^L_{j=1} a_{ij} y_{j,t-1} = a \beta^*_{1i} \Sum^L_{j=1} \gamma^*_{1j} y_{j,t-1}.
\end{equation}

The contagion parameters $a_{ij}$ can be decomposed: $a_{ij} = a \beta^*_{1i} \gamma^*_{1j}$, where a is a global contagion effect, $\beta^*_{1i}$ an index of vulnerability of segment $i$ to the infection and $\gamma^*_{1j}$ a measure of viral load of segment $j$. \footnote{The decomposition of $a_{ij}$ is  multiplicative, while an additive decomposition: $a_{ij} = \tilde{\alpha} + \tilde{\beta}_i + \tilde{\gamma}_j$, is used in the panel literature with two-ways fixed effects [see e.g. the running examples in Fernandez-Val, Weidner (2016) for probit and Poisson models]. In our framework it is not possible to transform the multiplicative form into an additive one by taking the logarithms of $a_{ij}, \beta^*_{1i}, \gamma^*_{1j}$, as there can exist individuals with zero value of $\beta^*_1$, such as the vaccinated or naturally immunized individuals. From the identification perspective, these zero values can be informative and should not be disregarded. Moreover, as noted in Chen et al. (2021), such interactive effects can capture network features as homophily and clustering.} Therefore the segments $i=1, \ldots, L$ can be ranked with  respect to their vulnerability  $\beta^*_{1i}$ and their viral load  $\gamma^*_{1i}$.

When $K$ is larger or equal to 2, a latent heterogeneity of segments arises with heterogeneity distribution $\pi$. Then, the segments can be ranked with respect to different notions of vulnerabilities, i.e. the $\beta^*_{ki}, k=1, \ldots, K$, and viral load, i.e. the $\gamma^*_{k,i}, k=1, \ldots, K$.

The potential interpretations of parameters $\beta^*, \gamma^*$, $\pi$ depend on the application of interest as shown in the example below.
\vspace{1em}

\textbf{Example 1~:} In the analysis of internet diffusion of messages (see Section 2.2.1), parameters $\beta^*$ (resp. $\gamma^*)$ can be used for ranking of receivers and senders, or followers and influencers.

\section{Identification of the Nonnegative Factorization}

The true NMF is not point-identified. This section discusses the identification issues and derives the identified set for small nonnegative ranks.

\subsection{The general framework}

 \setcounter{equation}{0}\def\theequation{3.\arabic{equation}}

The parametric model depends on the nonnegative matrix factorization~:

$$
A = BC' = (\beta_1, \ldots, \beta_K) (\gamma_1, \ldots, \gamma_K)' = \Sum^K_{k=1} \beta_k \gamma'_k,
$$

\noindent where $K$ denotes the nonnegative rank and $\beta_k, \gamma'_k s$ are the factorial directions. In practice, the structural parameters are $K, \beta_k, \gamma_k, k=1, \ldots, K,$ and there exists a large body of literature on the (lack of) identification of these parameters for a given matrix $A$.

It is easy to see that the factorization is not unique because the same matrix $A$ is obtained from a permutation of index $k$ and rescaling by a positive scalar, i.e. by replacing $\beta_k, \gamma_k$  by $\sigma_k \beta_k, \gamma_k/\sigma_k$ for a positive scalar $\sigma_k$. This identification issue is easily solved by a normalization [a free normalization in the terminology of Lewbel (2019), Section 6.3]. The more complicated identification issues that arise when $K \geq 2$ are discussed below under the following assumption~:\vspace{1em}

\textbf{Assumption A.3~:} The nonnegative rank of matrix $A$ is equal to the rank of $A$.\vspace{1em}

This assumption is not very stringent, even though some examples of nonnegative matrices with $Rk_+ (A) > Rk(A)$ have been given in the literature. It is useful to describe and parametrize the set of admissible NMF's when $Rk_+ (A) = Rk (A)$.

Assumption A.3 can be written under different equivalent forms~:

Assumption A.3 is satisfied

$\Longleftrightarrow$ the vectors $\beta_1, \ldots, \beta_K$ are linearly independent and

\hspace{1.5em} the vectors $\gamma_1, \ldots, \gamma_K$ are linearly independent\vspace{1em}

$\Longleftrightarrow$ $B'B$ and $C'C$ are invertible. \footnote{The NMF representation is different from the Singular Value Decomposition (SVD) of matrix $A$. In SVD the identification is usually solved by introducing the orthonormality restriction $B'B = C' C = Id$. Orthogonality is not possible in our framework since $\beta'_k \beta_l$ is always nonnegative.  Then the orthogonality condition would imply  $\beta_{ik} \beta_{il} = 0, \forall i$ and contradict assumption A.1.}\vspace{1em}

\noindent Assumption A.3 implies that the matrices $\beta_k \gamma'_l, k,l=1,\ldots,K$ are linearly independent as shown in Lemma 1 below :\vspace{1em}

\textbf{Lemma 1~:} Under Assumption A.3, $B \Delta C' = 0 \Rightarrow \Delta = 0.$\vspace{1em}

\textbf{Proof~:} Indeed $B\Delta C' = 0$ implies $B'B\Delta C'C=0,$ and then $\Delta = 0,$ since $B'B$ and $C'C$ are invertible.

\QED

Under assumptions A.1-A.3, the nonnegative rank is known as well as the rank of $A_0$. Let us first consider another factorization of matrix $A$ without taking into account the nonnegativity conditions on  $\beta_k, \gamma'_k s.$ Since the range of $A$ (resp. $A'$) is the space spanned by $\beta_1, \ldots, \beta_K$ (resp. by $\gamma_1, \ldots, \gamma_K)$, an alternative factorization is~:

$$
A = B\; G\; H' C' = B^* C^{*'},
$$

\noindent where $B^* = B\; G, C^* = C\;H$ and $G$, $H$ are invertible matrices \footnote{The columns of $B$ and the columns of $B^*$ are two bases of the range of $AA'$ equal to the range of $A$. Therefore they satisfy a one-to-one relationship represented by an invertible matrix $G$.}. Moreover, we have~:

$$
B\;G\;H'\;C' = B\;C',
$$

\noindent and, by Lemma 1, we deduce that $H' = G^{-1}$. Therefore we get~:

$$
B^* = BG, C^* = C (G')^{-1}.
$$

\noindent In addition, because of the definition of factors up to the permutation and (signed) scale effects, we can choose $G$ of the type~:

\begin{equation}
G = Q\; \mbox{diag}\; \sigma,
\end{equation}

\noindent where $\sigma = (\sigma_k), \sigma_k > 0$, and $Q$ a $(K,K)$ matrix with diagonal elements equal to 1. Then, $(G')^{-1} = (Q')^{-1} \; \mbox{diag}\; (1/\sigma)$.

Then, by taking into account the nonnegativity conditions, we get a constructive characterization of the identified set~:\vspace{1em}

\textbf{Proposition 1~:} For a specific factorization of matrix $A _0 : A_0=B_0 C'_0, B_0 \geq 0, C_0 \geq 0$ (referred to as a specific element of the identified set), all observationally equivalent nonnegative factorizations are such that~:

$$
B^* = B_0 Q \; \mbox{diag}\; \sigma,\; C^* = C_0(Q')^{-1} \; \mbox{diag}\; (1/\sigma), \;\sigma_k>0,  \forall k = 1,\ldots, K,
$$

\noindent where the matrix $Q$ is invertible, with unitary diagonal elements and such that~:

$$
B_0 Q \geq \;0,\; C_0 (Q')^{-1} \geq 0.
$$
\vspace{0.5em}

The nonnegative factorization is said to be essentially unique, or simply unique [see Laurberg et al. (2008)], if $Q=Id$ is the only solution to the set of inequalities given in Proposition 1.

\subsection{Special Cases}

\subsubsection{Case $K= Rk_+ (A) =1$}

When $K=1, A = \beta_1 \gamma_1', Q = (q_{11}) = (1)$, the nonnegative factorization is essentially unique. Moreover, we note that~:

$$
A \beta_1 = \beta_1 (\gamma'_1 \beta_1), A' \gamma_1 = \gamma_1 (\beta'_1 \gamma_1).
$$

\noindent It follows that $\beta_1$ (resp. $\gamma_1$) is an eigenvector of $A$ (resp. $A'$) associated with the eigenvalue $\gamma'_1 \beta_1 = \beta'_1 \gamma_1$, which is strictly positive. By the Perron-Froebenius Theorem [see  Meyer (2000)] the nonnegative matrix $A$ (resp. $A'$) has a unique eigenspace of dimension 1 generated by a nonnegative eigenvector, here $\beta_1$ (resp. $\gamma_1$). \footnote{It is easy to check that $\beta_1$ (resp. $\gamma_1$) is also an eigenvector of $AA'$(resp. $A' A$). Therefore they are elements of a singular value decomposition (SVD) (see Remark 3 in Section 4.1).}

\subsubsection{Case $K = Rk_+ (A)=2$}

For $K=2$, we get $Q = \left( \begin{array}{ll} 1 & q_{12} \\ q_{21} & 1\end{array}\right)$, and 

$$
(Q')^{-1} = \Frac{1}{1-q_{12} q_{21}} \left(\begin{array}{cc} 1 & -q_{21} \\ -q_{12} & 1\end{array} \right).
$$

\noindent Without the subscript $0$  for the "specific factorization of true matrix", the inequality restrictions in Proposition 1 become~:

$$
\left\{
\begin{array}{l}
  \beta_{1,j} + q_{21} \beta_{2,j} \geq 0,\; j=1,\ldots,n,\; q_{12} \beta_{1,j} + \beta_{2,j} \geq 0, \;j=1,\ldots, n, \\ \\
  \Frac{1}{1-q_{12} q_{21}} (\gamma_{1,j} - q_{12} \gamma_{2,j}) \geq 0, j=1,\ldots, m, \Frac{1}{1-q_{12} q_{12}} (-q_{21} \gamma_{1,j} + \gamma_{2,j}) \geq 0, \\ \\ 
  j=1,\ldots, m.
\end{array}
\right.
$$

\noindent It is easy to check  that these inequalities imply  $1-q_{12} q_{21} >0.$ Note also that an interesting case arises when we focus our attention on the  local identification in a neighbourhood of $(\beta_1, \beta_2), (\gamma_1, \gamma_2), $ i.e. in a neighbourhood of $q_{12} = q_{21} = 0$. Then, the system of inequalities is equivalent to~:

$$
\left\{
\begin{array}{l}
  \beta_{1,j} + q_{21} \beta_{2,j} \geq 0, \forall j,\; \mbox{with}\; \beta_{2,j} >0,\; -q_{21} \gamma_{1,j} + \gamma_{2,j} \geq 0, \forall j,\; \mbox{with}\; \gamma_{1,j} > 0,\\ \\
  q_{12} \beta_{1,j} + \beta_{2,j} \geq 0, \; \forall j,\; \mbox{with}\; \beta_{1,j} > 0,\; \gamma_{1,j} - q_{12} \gamma_{2,j} \geq 0, \forall j, \; \mbox{with}\; \gamma_{2,j} > 0.
\end{array}
\right.
$$

\noindent or

$$
\left\{
\begin{array}{l}
  q_{21} \geq \sup_{j:\beta_{2,j} > 0} (-\beta_{1,j} / \beta_{2,j}), \; q_{12} \leq \inf_{j:\gamma_{1,j} > 0} (\gamma_{2,j} / \gamma_{1,j}), \\ \\
  q_{12} \geq \sup_{j:\beta_{1,j} > 0} (-\beta_{2,j} / \beta_{1,j}), \; q_{21} \leq \inf_{j:\gamma_{2,j} > 0} (\gamma_{1,j} / \gamma_{2,j}).
\end{array}
\right.
$$

\noindent We deduce the following result~:\vspace{1em}

\textbf{Proposition 2~:} For $K=Rk_+ (A)=2$, the admissible matrices

 $Q= \left( \begin{array}{cc} 1 & q_{12} \\ q_{21} & 1 \end{array} \right)$ are such that~:

$$
\begin{array}{l}
- \inf_{j:\beta_{2,j} > 0} (\beta_{1,j}/ \beta_{2,j}) \leq q_{21} \leq \inf_{j:\gamma_{1,j} > 0} (\gamma_{2,j} / \gamma_{1,j}), \\ \\
 -  \inf_{j:\beta_{1,j} > 0} (\beta_{2,j}/ \beta_{1,j}) \leq q_{12} \leq \inf_{j:\gamma_{2,j} > 0} (\gamma_{1,j} / \gamma_{2,j}).
\end{array}
$$

\noindent Therefore, among the $2 (n+m)$ inequality restrictions in Proposition 1, only four are active and the remaining ones are redundant.

We deduce  the necessary and sufficient exclusion conditions for essential uniqueness [see e.g. Brie (2015)].\vspace{1em}

\textbf{Corollary 1~:} Under Assumption A.3 and for $K=Rk_+ (A)=2$, the nonnegative factorization is essentially unique if and only if there exists at least one index $j_1$ such that $\beta_{1,j_1} = 0, \beta_{2,j_1} > 0$, one index $j_2$ such that $\beta_{1,j_2} > 0, \beta_{2,j_2} = 0$, one index $j_3$ such that $\gamma_{1,j_3} = 0, \gamma_{2,j_3} > 0$ and one index $j_4$ such that $\gamma_{1,j_4}> 0, \gamma_{2,j_4} = 0.$\vspace{1em}

Without the "exclusion" restrictions of Corollary 1, and in particular under Assumption A.2 iii), there exists a multiplicity of admissible nonnegative factorizations that are easily deduced from one of them. 

Let us now discuss the degree of underidentification.
Since the identification issue due to the product of factorial directions by nonnegative scalars is solved in representation (2.13), we can focus on the identification of  $\pi_k, \beta^*_k, \gamma^*_k, k=1,2$. For K=2, the decomposition (2.13) after transformation  $a$ is:

$$
A = a [\tilde{\pi}_1 \tilde{\beta}^*_1 \tilde{\gamma}^{*'}_1 + \tilde{\pi}_2 \tilde{\beta}^*_2 \tilde{\gamma}^{*'}_2].
$$

\noindent It is easy to check (see online Appendix 1) that~:

$$
\begin{array}{l}
\tilde{\beta}^*_1 = p_1 \beta^*_1 + (1-p_1) \beta^*_2, \; \mbox{with}\; p_1 = \beta'_1 e/ (\beta'_1 e + q_{21} \beta'_2 e), \\ \\
\tilde{\beta}^*_2 = p_2 \beta^*_1 + (1-p_2) \beta^*_2, \; \mbox{with}\; p_2 = q_{12} \beta'_1 e/ (q_{12} \beta'_1 e + \beta'_2 e), \\ \\
\tilde{\gamma}^*_1 = p_3 \gamma^*_1 + (1-p_3) \gamma^*_2, \; \mbox{with}\; p_3 = \gamma'_1 e/ (\gamma'_1 e - q_{12} \gamma'_2 e), \\ \\
\tilde{\gamma}^*_2 = p_4 \gamma^*_1 + (1-p_4) \gamma^*_2, \; \mbox{with}\; p_4 = -q_{21} \gamma'_1 e/ (q_{21} \gamma'_1 e - \gamma'_2 e), \\ \\
\mbox{and}\\ \\
\tilde{\pi}_1/\tilde{\pi}_2 = (\beta'_1 e + q_{21} \beta'_2 e) (\gamma'_1 e -q_{12} \gamma'_2 e)/(q_{12} \beta'_1 e + \beta'_2 e) (-q_{21} \gamma'_1 e + \gamma'_2 e).
\end{array}
$$\vspace{1em}

\textbf{Corollary 2~:} For $Rk (A) = Rk_+ (A) = 2,$ the components $\pi_k, \beta^*_k, \gamma^*_k, k=1,2$ are not identifiable, with a degree of underidentification equal to 2.\vspace{1em}

The set of admissible decompositions (2.13) is described by means of $q_{12}, q_{21}$. The conditions on $q_{12}, q_{21}$ derived in Proposition 2 remain valid when $\beta, \gamma$ are replaced by $\beta^*, \gamma^*$.

\subsubsection{General Case}

For $K=2$, the identifiable set is described by two parameters $q_{12}, q_{21}$ satisfying $2(m+n)$ inequality restrictions. These restrictions are linear and Proposition 2 shows that only 4 of them are active.

In the general case, the identifiable set is parametrized by $K(K-1)$ parameters that are the off-diagonal elements of matrix $Q$. These parameters satisfy $ K(n+m)$ restrictions by Proposition 1. Therefore, the degree of underidentification increases rather quickly with the nonnegative rank and the subset of active restrictions cannot be derived analytically. To determine these restrictions providing a simplified definition of the identified set,  numerical algorithms are needed such as Active Set Sequential Quadratic Programming algorithms, some of them being available from Artelys Knitro.

A similar problem arises in sharp set identification for discrete choice models and treatment effects analysis. The main difference is \footnote{Another difference is the parametric assumption being used in our framework and nonparametric methods being used in the existing literature (see, Section 5).} that those inequality restrictions are linear as for $K = 2$ in Section 3.2.2.\footnote{See, the linear moment functions defining the inequalities in Chernozhukov et al. (2007), Section 2.1.} while in our framework they are nonlinear when $K \geq 3.$ Indeed, for $K=3$, we get~: $Q = \left( \begin{array}{ccc} 1 & q_{12} & q_{13} \\ q_{21} & 1 & q_{23} \\ q_{31} & q_{32} & 1 \end{array} \right)$, and up to the determinant, matrix $(Q')^{-1}$ has cofactor elements, such as $1-q_{23} q_{32}$ for instance, that are quadratic in $Q$ (more generally, they are polynomials of degree less or equal to $K-1$). Moreover, it is easy to see that the identifiable set is not convex.

\section{Statistical Inference}

 \setcounter{equation}{0}\def\theequation{4.\arabic{equation}}

A major challenge for statistical inference is the lack of identification of the true NMF. This identification issue can be addressed either by introducing identification restrictions and applying the standard maximum likelihood approach, or by estimating the set of all identifiable NMF directly. These two approaches are linked. Indeed, as shown in Section 3.3.2 for $Rk_+ (A) = 2$, the set of all identifiable NMF's can be deduced from one of them. Therefore, we can first identify one NMF and next deduce all the remaining ones from the identified one.

The proposed method proceeds as follows:  1) We show the convergence of the set of maximum likelihood estimators of $B, C$ to the identified set; 2) For a fixed number of observations, a multiplicity of ML estimators is obtained. Therefore, we introduce an alternating ML algorithm to fix the selected ML estimator for any $T$. This sequence of alternating ML estimators is well-defined, i.e. it converges to the identified set but not pointwise to a given element of this set; 3) We fix a benchmark in the identified set, which is the solution of an auxiliary optimization. That objective function is optimized with respect to an alternative parametrization in $(a, \pi, \beta^*, \gamma^*)$. This step provides a consistent estimator of a specific element of the identified set. 

The proposed method, called the identifying maximum likelihood (IML) approach is used to estimate the identified set. We derive its asymptotic distributional properties and show how to estimate the identified set and analyze its key properties. 

The approach is feasible because of the properties of the maximum likelihood estimator reviewed below.

\subsection{The ML approach}

\medskip

\subsubsection{Consistency}

Let us assume that the network model is well-specified and the true transition is~:

\begin{equation}
  l(y_t | y_{t-1} ; A_0) = l(y_t | y_{t-1}; B_0 C'_0),
\end{equation}

\noindent with a nonnegative rank $K_0$ of matrix $A_0$ assumed to be known. We consider a constrained ML maximization, providing:

\begin{equation}
  (\hat{B}_T,\hat{C}_T) = \arg \max_{B\geq0, C\geq0} \Sum^T_{t=1} \log l(y_t | y_{t-1}; B C').
\end{equation}

\noindent Due to the identification issue, there can be a large multiplicity of solutions to the finite sample optimization (4.2).
Under standard regularity conditions given in online Appendix 2, we can derive a consistency property of the above set of ML estimators.\vspace{1em}

\textbf{Proposition 3~:} The set of  ML estimators $(\hat{B}_T, \hat{C}_T)$ converges to the set  $\mathcal{A}_0$ of NMF associated with $A_0 = B_0 C'_0$ when $T$ tends to infinity.\vspace{1em}

More precisely let the Euclidian distance  denoted by
$d(.,.)$ on $\eR^{K(n+m)}$, $NMF_0 = \{(B,C), B \geq0, C\geq0$, with $BC' = B_0 C'_0\}$, and $\mathcal{D} [(\hat{B}_T, \hat{C}_T), NMF_0] = \min_{(B,C) \in NMF_0} d [(\hat{B}_T, \hat{C}_T), (B,C)]$, then $\mathcal{D}[(\hat{B}_T, \hat{C}_T), NMF_0$] tends to zero, when $T$ tends to infinity.
Under the regularity conditions, this convergence is uniform in $(\hat{B}_T, \hat{C}_T)$. It will also imply the convergence to this set of the well-defined alternating ML estimator introduced in the next section.

Thus, the $(\hat{B}_T, \hat{C}_T)$ does not necessarily converge to the true factorization $B_0, C_0$ due to the identification issue, but for a large $T, (\hat{B}_T, \hat{C}_T)$ it is close to another admissible NMF that can depend on $T$. \vspace{1em}

\subsubsection{Alternating ML (AML) Algorithm}

A ML estmator does not always have a closed-form. Hence, in practice it is computed numerically from an algorithm, such as a Newton-Ralphson type of algorithm. In our framework of partial identification, a Newton-Ralphson type of algorithm cannot be used jointly for $B$ and $C$. The reason is that each iteration requires the inversion of a Hessian matrix that is not invertible due to the identification issue. The AML algorithm solves this issue.

In the presence of a multiplicity of NMFs, we apply an alternating AML algorithm [Gourieroux, Monfort, Renault (1990), Kim, Park (2007)]. We observe that even if the factorization $B,C$ is not identifiable, $B$ (resp. $C$) is identifiable when $C$ is known (resp. $B$ is known). This leads to the following alternating AML algorithm, where at step $p,$ $\hat{B}_{p,T}, \hat{C}_{pT}$ is computed and then $\hat{B}_{p+1,T}, \hat{C}_{p+1,T}$ are obtained as follows:

\begin{eqnarray}
  \hat{B}_{p+1,T} &=& \arg \max_{B \geq 0} \Sum^T_{t=1} \log l (y_t | y_{t-1} ; B \hat{C}'_{p,T}), \\
  \hat{C}_{p+1,T} &=& \arg \max_{C \geq 0} \Sum^T_{t=1} \log l (y_t | y_{t-1} ; \hat{B}_{p+1,T} C').
\end{eqnarray}

\noindent By construction, this algorithm produces at each  iteration $p$ a higher value of the log-likelihood function than at iteration $p-1$.

We have to distinguish the ML estimator from the alternating ML estimator obtained from the algorithm (4.3)-(4.4). As mentioned earlier, when some parameters are not identifiable there is a multiplicity of ML estimators. However, there is a unique sequence of alternating AML estimators for the given starting values, even though the AML algorithm does not necessarily numerically converge pointwise, due to the identification issue.

\subsection{Nonnegative rank $K_0 = Rk_+ (A_0) = 1$}

As pointed out in Section 3.2.1, the NMF is essentially unique for $K_0 = 1.$ This greatly simplifies the estimation and explains its use in applied econometrics. Because the ML estimator is unique in this case, it can be computed from a standard Newton-Ralphson algorithm.

We introduce the identification restrictions: $A = a \beta \gamma'$, with $a=0, \beta \geq 0, \gamma \geq 0$ and $\beta' e = \gamma' e = 1.$ Let us now consider the dynamic model described in Section 2.2.3. and based on the latent parametric model $l (y;\theta)$, where $\theta$ is replaced by $Ay_{t-1}$ [see, Section 2.2.3]. We get~:

\begin{equation}
  l(y_t | y_{t-1} ; A) = l (y_t ; a \beta \gamma' y_{t-1}).
\end{equation}

\noindent The partial derivatives of the log-likelihood with respect to $a, \beta, \gamma$ are easily derived from the partial derivatives of the latent log-likelihood with respect to $\theta$. We have~:

\begin{equation}
\left\{
\begin{array}{lcl}
\Frac{\partial \log l}{\partial \beta} (y_t | y_{t-1} ; A) & = & a \gamma' y_{t-1} \Frac{\partial \log l}{\partial \theta} (y_t ; a \beta \gamma' y_{t-1}), \\ \\
\Frac{\partial \log l}{\partial \gamma} (y_t | y_{t-1} ; A) & = & a y_{t-1} \beta' \Frac{\partial \log l}{\partial \theta} (y_t ; a \beta \gamma' y_{t-1}), \\ \\
\Frac{\partial \log l}{\partial a} (y_t | y_{t-1} ; A) & = &  \gamma' y_{t-1} \beta' \Frac{\partial \log l}{\partial \theta} (y_t ; a \beta \gamma' y_{t-1}).
\end{array}
\right.
\end{equation}

\noindent The asymptotic properties, especially the asymptotic distribution of the ML estimators of $a, \beta, \gamma,$ depend on the location of the true vectors $\beta_0, \gamma_0$. These properties are straightforward under the assumption A.2 iii) below.\vspace{1em}

\textbf{Positivity Assumption A.2 iii):} The entries of $\beta_0$ and $\gamma_0$ are strictly positive.\vspace{1em}

Under this positivity assumption, the non-negativity-constrained ML estimators have asymptotically strictly positive entries, and the unconstrained and non-negativity constrained estimators are asymptotically equivalent. However, the ML estimator has to account for the linear constraint of unit sum. This estimator without the non-negativity restrictions is defined as~: 

$$
\begin{array}{rcl}
(\hat{a}, \hat{\beta}, \hat{\gamma}) & =& \arg \max_{a,\beta, \gamma} \Sum^T_{t=1} \log l (y_t;a \beta \gamma' y_{t-1}), \\ \\
\mbox{s.t :}\; \beta' e & = & \gamma' e = 1.
\end{array}
$$

\noindent The first-order conditions for the Lagrange multipliers associated with the linear restrictions and denoted by $\lambda, \mu$ ,  are~:

$$
\begin{array}{l}
  \Sum^T_{t=1} [a \gamma' y_{t-1} \Frac{\partial \log l}{\partial \theta} (y_t; a \beta \gamma' y_{t-1})] - \lambda = 0, \\ \\
   \Sum^T_{t=1} [a  y_{t-1} \beta'  \Frac{\partial \log l}{\partial \theta} (y_t; a \beta \gamma' y_{t-1})] - \mu = 0, \\ \\
    \Sum^T_{t=1} [ \gamma' y_{t-1} \beta' \Frac{\partial \log l}{\partial \theta} (y_t; a \beta \gamma' y_{t-1})]  = 0,\\ \\
    \beta' e = \gamma' e = 1.
\end{array}
$$

\noindent The FOC need to be solved in $a, \beta, \gamma, \lambda, \mu$.\vspace{1em}

Asymptotically, when $T$ tends to infinity, we get consistent and asymptotically normal ML estimators of $B$ and $C$. Their asymptotic variance-covariance matrix has a standard form [see Gourieroux, Monfort (1995), Section 10.3] and its estimate can be computed by the software.\vspace{1em}

 \textbf{Remark 3~:} As pointed out in Section 3.2.1, $\beta_1$ (resp. $\gamma_1)$ can be interpreted as an eigenvector of $A$ (resp. $A'$). It is easy to check that $\beta_1$ (resp. $\gamma_1$) is an eigenvector of $AA'$ (resp. $A'A$). This implies that $A=\beta_1 \gamma'_1$ is  a singular value decomposition (SVD) of matrix $A$, for which statistical inference is available mainly in a Gaussian framework [Anderson, Rubin (1956), Anderson (1963), Tipping, Bishop (1999)].

 However, for $Rk (A_0) = Rk_+ (A_0) = 1$, the standard SVD estimation method is not relevant, as it does not account for the  non-negativity of the data and the non-negativity of matrix $A_0$. Moreover, the SVD interpretation of the NMF is no longer valid for $Rk_+ (A_0) \geq 2$. Indeed, matrix $AA'$ (resp. $A'A$) is also non-negative, and, except the Perron, Froebenius eigenvector $\beta_1$ of $AA'$, all other eigenvectors $\beta_2, \beta_3$ must have at least one negative, or non-real component.

\subsection{Nonnegative rank $K_0 = Rk_+ (A_0) \geq 2.$}

As mentioned in Section 3.2.3, it is sufficient to estimate one of the admissible NMF's to deduce  the sharp identified set. The problem with applying the AML method is that the convergence of the AML estimator to the identifed set does not imply its convergence to a given NMF.

This section shows how an  IML algorithm can solve this issue, allowing us to derive the asymptotic distribution of the set of admissible NMF's. For expository purpose, we provide in the text the assumptions specific to our problem.  The additional assumptions needed for asymptotic analysis are given in online Appendix 2.

\subsubsection{Consistency of the alternating ML estimator}

In this section we consider the consistency of the AML approximation to the identified set when $T$ tends to infinity and the number of iterations $p_T$ in the AML algorithm depends on T in a suitable manner.

Let us consider the dynamic model~:

\begin{equation}
  l(y_t | y_{t-1}; A) = l(y_t; a \Sum^K_{k=1} \pi_k \beta^*_k \gamma^{*'}_k y_{t-1}),
\end{equation}

\noindent with $\beta^{*'}_k e = \gamma^{*'}_k e = 1, k=1,\ldots, K, \;\pi' e = 1$ and the identified set:

$$
\mathcal{A}_0 = \{ a, \pi_k, \beta^*_k, \gamma^*_k, k=1,\ldots, K,\; \mbox{such that}\; a \Sum^K_{k=1} \pi_k \beta^*_k \gamma^{*'}_k = A_o\}.
$$

\noindent where $A_0$ is the true value of $A$.

Section 2.2 shows that the underlying parametric families $l(y;\theta), \theta \geq 0$, are often constructed from the products of Poisson, or exponential distributions. Therefore, they satisfy the following assumption~:\vspace{1em}

\textbf{Assumption A.4~:} The underlying log-likelihood $\log l (y;\theta)$ is concave in $\theta, \theta \geq 0.$\vspace{1em}

Under Assumption A.1 and for any value of the number of observations $T$ , each step of the  AML algorithm outlined in Section 4.1.2 leads to a unique solution in $(B,C)$,  because the objective function is log-concave in $B$ (resp. $C$) for a given $C$ (resp. $B$), and also under the alternative parametrization $(\pi_k, \beta^*_k, \gamma^*_k, k=1,\ldots, K).$ Therefore, the AML estimator is a function of the underlying (normalized) log-likelihood $\Frac{1}{T} L_T (.) = \Frac{1}{T} \Sum^T_{t=1} \log l(y_t ; .),$ and of the initial values used in the algorithm (and of the number of iterations $p$). Let the set of parameters be denoted by  $\alpha = (\pi_k, \beta^*_k, \gamma^*_k, k=1,\ldots, K)$ and the selected initial value by $\alpha^o $. The AML estimator at iteration $p$ can be written as~:

\begin{equation}
 \hat{\alpha}_T (\alpha^o, p) = m (\Frac{1}{T} L_T (.) ; \alpha^o, p),
\end{equation}

\noindent where $m$ is a deterministic function. Then, for large $T$, the AML estimator will converge asymptotically to the value~:

\begin{equation}
  \alpha_\infty (\alpha^o, p) = m [E_0 \log l (Y_t; A_0); \alpha^o, p],
\end{equation}

\noindent that belongs in the set $\mathcal{A}_0$ if $p$ is large. Note, that this limiting value can depend on the initial value $\alpha^0$ used in the algorithm.\vspace{1em}

\noindent More precisely, under Assumption A.1-A.4 and the additional regularity conditions a.1 given in online Appendix 2, we have the following proposition that is a direct consequence of the numerical consistency of the AML approximation (for $p \rightarrow \infty$ and $T$ fixed) and of the uniform convergence in Proposition 3~:\vspace{1em}

\textbf{Proposition~4: } For large $T$, there exist a function $c(.)$ and a number of iterations $p_T$ such that, for any $p\geq p_T$~:

$$
\mathcal{D} [\alpha_T (\alpha^o, p), \mathcal{A}_o] < c(\alpha^o)/T,
$$

\noindent where $\mathcal{D}$ measures the distance between $\alpha_T (\alpha^o, p)$ and the set $\mathcal{A}_o.$\vspace{1em}

\noindent In practice, we can apply the AML algorithm with a given starting value $\alpha^o$ and number of iterations $p$. The number $p$ needs to be set sufficiently large for Proposition 4 to be satisfied. Then, the asymptotic bias of the alternating ML estimator will be sufficiently small to be negligible in the asymptotic distribution of the IML estimator derived later in the text. Moreover, two optimizations are performed at each step of the algorithm, with respect to $B$ and $C$, respectively. A Newton-Ralphson type of algorithm can be used in this case. Under the log-concavity assumption A.4, the Newton-Ralphson algorithm is a special case of the steepest ascent algorithm, which increases the objective function at each step. As the increase of the objective function at each step of the algorithm is ensured, the IML with a fixed number of iterations for the intermediate optimization will also increase the objective function, that is sufficient for the numerical consistency of the AML algorithm under Assumption A.4.

\noindent It follows from Proposition 1 that all other elements of $\mathcal{A}_0$ are functions of $\alpha_\infty (\alpha^o; p)$ and the elements of a matrix $Q$ with unitary diagonal elements are such that~:

$$
B[ \alpha_\infty (\alpha^o, p)] Q \geq 0, \; \;  C[ \alpha_\infty (\alpha^o, p)] [Q']^{-1} \geq 0,
$$

\noindent and define a set $\mathcal{Q} [\alpha_\infty (\alpha^o, p)]$ of admissible values of transformation $Q$.

\noindent Equivalently, we have a parametric representation of the set $\mathcal{A}_0$ of NMF's~:

\begin{equation}
  \mathcal{A}_0 = \{ \alpha : \alpha = \xi [\alpha_\infty (\alpha^o, p), Q], Q \in \mathcal{Q} [\alpha_\infty (\alpha^o,p)]\},
\end{equation}

\noindent where $\xi$ is a known function.\vspace{1em}

\noindent Then, the set $\mathcal{A}$ is consistently estimated as:

\begin{equation}
  \mathcal{\hat{A}}_T = \{ \alpha : \alpha = \xi [\hat{\alpha}_T (\alpha^o, p), Q], Q \in \mathcal{Q} [\hat{\alpha}_T (\alpha^o, p)] \equiv \hat{Q}_T (\alpha^o, p)\}.
\end{equation}

The estimation method presented above will allow us to approximate a possibly $p$-dependent element of the identified set for a sufficiently large $p$ .

Proposition 4 implies that the above AML estimator is a maximizer of the log-likelihood function. This AML estimator can be used to derive a Monte Carlo estimator of the identified set based on a quasi-likelihood ratio with the quantiles computed by simulations [see, Chen, Christensen, Tamer (2018), Remark 1, p. 1972]. In our framework, the chosen quasi-prior distribution would need to ensure the existence of the log-likelihood function.  For example, it has to satisfy the non-negativity restrictions on matrix $A$ in a Poisson autoregressive model. Moreover, the generic form of the estimated confidence set, i.e. $\{ A: L_T(A) \geq \xi \}$, where $L_T$ denotes the log-likelihood and $\xi$ is the estimated quantile, is hard to determine unless the appropriate parametrization given in Proposition 1 is used.

Our objective is to find a consistent ML estimator of the identified set, derive the asymptotic distribution of the estimated identified set
and of the elements in that set, and find the rank-constrained ML estimator of $A$ that maximizes the log-likelihood function under the constraint $A=BC'$, and the associated asymptotic efficiency bound. Because of the lack of identification, we do not have the numerical stability of 
$\hat{\alpha}_T (\alpha_0,p)$, for large $p$. Therefore, we cannot expect to prove any asymptotic normality of this AML estimator. In order to stabilize the algorithm, we include an additional optimization step. 

\subsubsection{How to introduce an identification restriction by Identifying Maximum Likelihood}

An identification issue is commonly solved either by introducing implicit identification restrictions, or by reparametrizing the model and dividing the parameters into the set of identifiable and non identifiable parameters in an appropriate way (this is the "global" reduced form reparametrization considered in Chen et al. (2018), Section 5.1.1). These approaches are not suitable for our framework, where the parametrization of the identified set depends on a selected element of $\mathcal{A}_0$. Hence, we introduce indirectly $K(K-1)$ identification restrictions. 

Let us consider an alternative parametrization method: When a specific element of the identified set is known, the identifiable set is parametrized by $vec^* Q=q$, where $vec^* Q$ denote the stacked elements of $Q$ except for the diagonal elements equal to 1. This parametrization cannot be used, as long as that specific element is unknown.
However, when a specific element $\tilde{\alpha}$ say, is given, it defines a new origin and then a new parametrization by $q$ of the identifiable set is obtained.

Let us now determine a benchmark $\alpha_0^*$. We define a benchmark element of the identified set as the  optimizer of an additional criterion with respect to the additional parameter $q$. Two criteria $\tilde{g} (q,\tilde{\alpha}) = g(\alpha),$ where $\alpha \in \mathcal{A}_0$ arise naturally~:\vspace{1em}

i) The concentration of a discrete probability distribution is usually measured by $\Sum^K_{k=1} (\pi_k \log \pi_k)$. This quantity is negative and it is minimized for $\pi_k = 1/K, \forall k,$ that is a uniform distribution, and it increases to zero with the concentration of the distribution. Therefore, we can choose as the benchmark specific element, the factorization providing the most concentrated latent heterogeneity defined by~:

$$
\alpha^*_0 = \arg \max_{\alpha \in \mathcal{A}_0} \Sum^K_{k=1} (\pi_k \log \pi_k).
$$

ii) An alternative criterion is~:

$$
g(\alpha) = \det (\tilde{B}' \tilde{B}),\:\mbox{where}\; \tilde{B} = (\beta^*_1, \ldots, \beta^*_K),
$$

\noindent or $g(\alpha) = \det (\tilde{C}' \tilde{C}),$ where $\tilde{C} = (\gamma^*_1, \ldots, \gamma^*_K)$.\vspace{1em}

The above criterion measures the volume of the parallelepiped generated by the columns of $\tilde{B}$ (resp. $\tilde{C})$ [see e.g. Barth (1999)]. The larger this volume, the less "colinear" the columns of $\tilde{B}$ (resp. $\tilde{C}$). \footnote{These criteria are the analogues of the identification restrictions introduced for SVD: $B'B = C'C = Id$, where all factorial directions are orthonormal.}
These criteria could be used jointly.\vspace{1em}

To proceed with the algorithm-based identification of the benchmark $\alpha_0^*$ by means of this additional optimization, an additional step needs to be
included in the AML algorithm.\footnote{This additional step is the analogue of step 2 in the estimation approach introduced in Davezies et al. (2022), Section 3.1., where it is applied to an approximate identified set instead of the identified set itself.} Let 
the recursive system in this algorithm be denoted by:

$$
\alpha^{(p+1)} = H (\alpha^{(p)}),
$$

\noindent where $H$ depends on the observations.\vspace{1em}

\noindent Then, the identifying maximum likelihood (IML) algorithm is the following~:\vspace{1em}

\textbf{step 1:} Select an initial value $\alpha^{(0)}$.

\textbf{step $p$ :} At step $p$, a value $\alpha^{(p)}$ is available.

i) Apply the AML algorithm to get a value $\tilde{\alpha}^{(p+1)} = H (\alpha^{(p)})$.

This value is considered as an approximation of a point in the identifiable set.

It can be used to parametrize the set $\hat{\mathcal{A}}^{(p)}$ by $q$.

\medskip

ii) Perform the optimisation of the additional criterion to get~:

$$
q^{(p+1)} = \mbox{Opt}_{q \in Q^{(p+1)}} \tilde{g} (q; \tilde{\alpha}^{(p+1)}),
$$

\noindent where $Q^{(p+1)}$ is the domain defined by the inequality restrictions applied with $\tilde{\alpha}^{(p+1)}$. Then the solution is a function of $\tilde{\alpha}^{(p+1)}$, that is~: $ q^{(p+1)} = q(\tilde{\alpha}^{(p+1)}),$ say, where $q(.)$ does not depend on the observations.

\medskip

iii) Find $\alpha^{(p+1)}$ by transforming $\tilde{\alpha}^{(p+1)}$ with a linear transformation $Q^{(p+1)}$ associated with $q^{(p+1)}$, including this choice of decreasing ordering of index to get $\pi_k^{(p+1)}$.
More precisely:

compute $Q^{(p+1)}$ such that $q{(p+1)} = vec^* Q^{(p+1)}$,

compute $\tilde{B}^{(p+1)}, \tilde{C}^{(p+1)}$ from $\tilde{\alpha}^{(p+1)}$

compute $B^{(p+1)} = \tilde{B}^{(p+1)} Q^{(p+1)}$, $C^{(p+1)} = \tilde{C}^{(p+1)} [Q^{(p+1)'}]^{-1}$

compute $\alpha^{(p+1)}$ from $B^{(p+1)}, C^{(p+1)}$, etc. 

\medskip

The main difference between the AML and IML algorithms concerns the consistency. The AML converges to the set 
$\mathcal{A}_0$, but the convergence is not pointwise. The IML converges to the given $\alpha_0^*$ that allows us to perform a Taylor expansion of the first-order conditions in order to derive the asymptotic normality.

The IML method requires the availability of algorithms for the optimization of nonlinear functions under a large number of nonlinear inequality restrictions. The recent developments in Active Set Sequential Quadratic Programming (SQP) have largely solved this problem [see e.g. Gill et al. (2002), Nocedal, Wright (2006), and Liu (2005) for a proof of numerical convergence].\vspace{1em}

\textbf{Remark 4~:} The additional intermediate optimization in the IML algorithm does not necessarily have  to be introduced starting from the first iteration $p=1$. It can be introduced later, when $p$ is sufficiently large to get a value close to the identified set, by Proposition 4. In this respect, the IML is used to stabilize pointwise the values obtained from a standard ML algorithm.

\subsubsection{Asymptotic distributions}

The benchmark $\alpha^*_0$ can be either in the interior of the identifiable set, or on its boundary. The asymptotic normality cannot be expected in the latter case, but standard asymptotic arguments can be used to derive the asymptotic normality of the IML estimator adjusted to reach $\alpha^*_0,$ if $\alpha^*_0$ is in the interior of $\mathcal{A}_0$ and the associated $\pi_{0k}^*$ are all distinct \footnote{This additional condition is analogous to the condition of distinct eigenvalues in the joint spectral decomposition of $AA'$ and $A'A$ in the standard SVD.}.



To clarify the role of the intermediate optimization in the IML algorithm, let us first consider the standard information matrices. Two matrices appear naturally\footnote{For expository purpose, we keep the same notation $l$ for the conditional likelihood as a function of $A$, or a function of $\alpha$.}~:

i) The unconstrained information based on $A$, is:

$$
E_0 \left[ -\Frac{\partial^2 \log l(y_t|y_{t-1};A)}{\partial \;\mbox{vec}\;
  A\; \partial \;\mbox{vec}\;A'}\right].
$$

This matrix is invertible by the assumption of identifiable $A$, but can be of a high dimension in practice.

ii) The information matrix corresponding to parameter $\alpha$ is: \linebreak
$J_0 = E_0 [- \Frac{\partial^2 \log l(y_t|y_{t-1};\alpha)}{\partial \alpha \partial \alpha'}]
$ constrained by the unit mass restrictions on $\pi, \beta^*_k, \gamma^*_k, k=1,\ldots, K$.

This matrix has a smaller dimension, but is not of full rank due to the lack of identification.\vspace{1em}

The algorithm introduced above has extended the constrained maximum likelihood approach by adding asymptotically the identification restrictions corresponding to the first-order conditions of the optimization of $\tilde{g} (q, \alpha)$ with respect to $q$, that are~:

\begin{equation}
  \Frac{\partial \tilde{g} (q, \alpha)}{\partial q} = 0 \Rightarrow q = q(\alpha),
\end{equation}

\noindent of a number equal to the degree of underidentification.

These limiting conditions have been replaced by $\Frac{\partial \tilde{g}}{\partial q} [q (\hat{\alpha}_T), \hat{\alpha}_T] = 0$ in the IML algorithm and can be expanded in a neighbourhood of $[q (\alpha^*_0) = 0, \alpha^*_0]$. We get~:

$$
\left[\Frac{\partial^2 \tilde{g}}{\partial q \partial q'} [0, \alpha^*_0] \Frac{dq}{d\alpha} (\alpha^*_0) + \Frac{\partial^2 \tilde{g}}{\partial q \partial \alpha'}(0,\alpha^*_0)\right] \sqrt{T} (\hat{\alpha}_T - \alpha^*_0) \simeq 0,
$$

\noindent that are the additional linear restrictions $D'_2 \sqrt{T} (\hat{\alpha}_T - \alpha^*_0) = 0$ on $\sqrt{T} (\hat{\alpha}_T - \alpha^*_0)$  to be taken into account for the computation of the asymptotic variance of $\hat{\alpha}_T$. More precisely,\vspace{1em}

\textbf{Proposition 5~:} If $\alpha^*_0$ is in the interior of $\mathcal{A}_0$, if the associated $\pi_{0k}^*$ are all distinct, and if assumptions A.1-A.4 and the additional regularity assumptions a.1-a.2 are satisfied, then~;

$$
\sqrt{T} (\hat{\alpha}_T - \alpha^*_0) \stackrel{d}{\longrightarrow} N (0, J^{11}_0 \; J\; J^{11}_0),
$$

\noindent where $J^{11}_0$ is the North-West block in the block decomposition of the inverse of matrix $\left( \begin{array}{cc} J_0 & D \\ D' & 0\end{array}\right), D = (D_1, D_2), D_1$ defining the $2K+1$ unit mass restrictions on $\pi_k, \beta^*_k, \gamma^*_k, k=1,\ldots, K$, and $D_2$ the linearized restrictions corresponding to the intermediate optimizations in the IML algorithm.

\medskip
In the expression of the asymptotic variance-covariance matrix, we observe the three main components of the information related with the unconstrained ML of $\alpha$, the unit mass restrictions $D_1$ and the restrictions $D_2$ due to additional intermediate optimisation.\vspace{1em}

\textbf{Proof~:} i) The proof is standard and based on the asymptotic expansion of the first-order conditions on the Lagrangean to account for the equality restrictions (note that the inequality restrictions are not binding if $\alpha^*_0$ belongs in the interior of $\mathcal{A}_0)$. These asymptotic expansions are~:

 \begin{equation}
\left( \begin{array}{cc} J_0 & D \\ D' & 0 \end{array}\right)
 \left[ \begin{array}{l} \sqrt{T} (\hat{\alpha}_T - \alpha^*_0) \\ \sqrt{T} (\hat{\lambda}_T - \lambda^*_0)\end{array} \right]
 \simeq
 \left[ \begin{array}{c}\Frac{1}{\sqrt{T}} \Sum^T_{t=1} \Frac{\partial \log l}{\partial \alpha} (y_t|y_{t-1}; \alpha^*_0)\\ 0 \end{array}\right],
\end{equation}

\noindent where $\hat{\lambda}_T$ is the associated estimator of the Lagrange multipliers. Because $\alpha_0^*$ is a maximizer of $E_0 \log l(y_t | y_{t-1}, \alpha)$,
the normalized score in (4.13) is asymptotically normally distributed with mean zero and variance $J$.

\vspace{1em}

The result follows whenever the matrix $\left( \begin{array}{cc} J_0 & D \\ D' & 0 \end{array}\right)$ is invertible. \footnote{We cannot use the usual block formula to compute $J^{11}_0$ [see e.g. Gourieroux, Monfort, Section 10.3.b) because $J_0$ is not invertible due to the identification issue. However, it is easy to check that a closed form expression of the asymptotic variance of the estimator is~:

$$
[(Id-P) J_0 (Id-P)+P]^{-1} (Id-P) J_0 (Id-P) [(Id-P) J_0 (Id-P) + P]^{-1},
$$.}

ii) Let us now discuss this invertibility condition by finding the null space of this matrix, i.e.  the solutions  $\theta, \lambda$ of the system~:

$$
\left\{
\begin{array}{l}
J_0 \theta + D \lambda = 0, \\
D' \theta = 0.
\end{array}
\right.
$$

\noindent We know that  $D' J_0 \theta + D' D \lambda = 0 \Rightarrow \lambda = - (D' D)^{-1} D' J_0 \theta.$ Then the system in $\theta$ only is~:

$$
\left\{
\begin{array}{l}
(Id-P) J_0 \theta = 0, \\
D' \theta = 0.
\end{array}
\right.
$$

\noindent where $P$ is the orthogonal projector on the space generated by $D$.

Since the columns of $(Id-P) J_0$ are orthogonal to the columns of $D$, we see that $\theta = 0$ is the unique solution if and only if~: $Rk ((Id-P) J_0) = \dim \alpha - \dim q -1 -2K$. This statement is Assumption a.2 viii) in online Appendix 2.

\QED

\textbf{Remark~5:} 
The sequence of optimizations cannot be replaced by a penalty term in the objective function.
By analogy with the $l^1$-penalty introduced in LASSO [Tibshirani  (1996)], the machine learning literature suggests to use a penalty in the objective function to ensure the (numerical) convergence of the ML algorithm [see e.g. Kim, Park (2008), Schachtner et al. (2011)], or to circumvent the curse of dimensionality [DePaula et al. (2020), eq. (10). See also Uhlig (2005), Appendix B.2 for applications to macro-economics]. In our framework, this approach would lead to be an objective function of the form $\log l_T (y;\alpha) + \lambda_T g (\alpha)$, where the tuning parameter $\lambda_T$ would be an appropriately chosen function of $T$ to ensure the numerical convergence. It is easy to see why this approach would not provide an estimator converging to a benchmark  element of the identified set: The asymptotic first-order conditions involve $\Frac{\partial \log l_T (y, \alpha)}{\partial \alpha} + \lambda_T \Frac{\partial g(\alpha)}{\partial \alpha}.$ They are not aligned with the direction allowing to remain in the set, since $\Frac{\partial g(\alpha)}{\partial \alpha} = \Frac{d\tilde{g} [q(\alpha), \alpha]}{d\alpha}$ differs from $\Frac{\partial \tilde{g} (q,\alpha)}{\partial q}.$
\vspace{1em}

The asymptotic Gaussian uncertainty is driving all the uncertainties on the true set $\mathcal{A}_.$ of NMF's. More precisely, any other element of $\mathcal{A}_0$ can be written as $\xi (q,\alpha^*_0)$, with $q \in Q(\alpha^*_0).$ Then, that element can be estimated by $\xi (q, \hat{\alpha}_T)$, which is a given function of $\alpha_T$. Therefore, it inherits the asymptotic properties of $\hat{\alpha}_T$: it is consistent of $\xi (q,\alpha^*_0)$, asymptotically normal, and its asymptotic variance-covariance matrix is obtained from the Slutsky formula, whenever $q$ is not on the boundary of $Q (\alpha^*_0)$. If $q$ is on the boundary, its distribution will become truncated normal, and can be easily found by simulation.\vspace{1em}

\textbf{Remark 6~:} The set $\mathcal{A}_0$ is of a large dimension and it is not possible to represent it in a 2 or 3-dimensional space. However it is possible to consider "cuts" of that set obtained by varying a given component of $q$ and setting the other components equal to zero, to examine how the NMF responds to changes in that component. This analysis will depend on the application. Let us consider the static NMF applied to image analysis. The criterion $\det (B' B)$ is likely a measure of contrast and the above approach could be used to find the component $q_j$ that is preferred for changing the contrast of the photo $BC'$ from low to high. Another direction could by used to manage the brightness of the image and so on.\vspace{1em}

For illustration, let us consider the matrix~: $A = \Frac{1}{10} \left( \begin{array}{ccc} 2&1&1\\ 1 & 1&1 \\ 1 & 1&1\end{array}\right)$. Its elements sum up to one and $A$ defines a joint probability distribution. It is easy to check that $Rk (A) = Rk_+ (A) = 2$. This matrix is a mixture of two joint distributions satisfying the independence condition, which can be written in an infinite number of ways. For example,  matrix $A$ can  be written as~:

$$A = \Frac{1}{4} \left( \begin{array}{c}1 \\ 0 \\0 \end{array}\right) (3/5, 1/5, 1/5)+ \Frac{3}{4} \left( \begin{array}{c}1/5 \\ 2/5 \\2/5 \end{array}\right) (1/3, 1/3, 1/3),
$$

\noindent or as~:

$$
A = \Frac{1}{10} \left( \begin{array}{c}1 \\ 0 \\0 \end{array}\right) (1,0,0)+ \Frac{9}{10}  \left( \begin{array}{c}1/3 \\ 1/3 \\1/3 \end{array}\right) (1/3, 1/3, 1/3).
$$

The volumes of the parallelepiped generated by the $\beta^*$ (resp. $\gamma^*)$ are $8/25$ for the first decomposition and $2/9$ for the second decomposition (resp. $8/225$ and $2/9$). The heterogeneity distribution is less concentrated in the first decomposition than in the second one, its $\beta^*$ vectors are less colinear and its $\gamma^*$ vectors are more colinear than in the second one. Note that for $K=2, \pi_1$ is an homographic function of $q_{12}$ for  a given $q_{21}$ (resp. of $q_{21}$ for a given $q_{12})$ and hence it is monotonic.\vspace{1em}

\textbf{Remark 7~:} The IML approach can be used to derive the lower and upper bounds for partially identified scalar parameters, and to obtain the measures of uncertainty on these bounds [see e.g. Imbens, Manski (2004), Stoye (2009)]. Typical examples are the minimum and maximum values of functions $\Sum^K_{k=1} (\pi_k \log \pi_k) \det (\tilde{B}' \tilde{B}), \det (\tilde{C}' \tilde{C}).$ This procedure is analogous to determining the confidence intervals for average marginal effects in a fixed effects panel logit model [Liu et al. (2021), Davezies et al. (2022)]. Likely, there exists an interval of admissible values of the above uncertainty measures. Such an interval is easy to obtain for the measure of concentration when $K=2$. However, the lower and upper bounds will be reached on the boundaries of the identified set, for example on the bounds for $q_{12}, q_{21}$ when $K=2$. Then, the joint asymptotic distribution of these bounds cannot be Gaussian due to the effect of infimum in Proposition 2.2. Therefore Assumption A.1 i) of normality of the upper and lower bounds in Imbens et al. (20045), Stoye (2009) is not satisfied in our framework. Note, that the joint asymptotic distributions of these two bounds are easily derived by simulations,  and  that, by construction,  we cannot have bound reversal in the estimation.\vspace{1em}

\subsubsection{Asymptotic distribution of $\hat{A}_T$}

The IML approach also helps find the estimates and confidence intervals of identifiable parameters, especially of the elements $a_{ij}$ of matrix $A$. In practice, we usually encounter the curse of dimensionality in the unconstrained estimation of $A$. Moreover, the estimator has to be applied under the constraint of a given non-negative rank~: $Rk_+ (A) = K$, and the rank-constrained confidence intervals are likely narrower than the unconstrained ones. They can be derived from $\hat{\alpha}_T$ by simulations, given that~:

$$
a_{ij} \simeq \hat{a} \Sum^K_{k=1} \hat{\pi}_k \hat{\beta}^*_{ik} \hat{\gamma}^*_{jk} = \hat{a}_{ij}.
$$

Asymptotically, $\hat{a}_{ij}$ will converge to the true value $a_{ij,0}$ that does not depend on the choice of $\alpha^*_0$ benchmark. Similarly, its asymptotic variance-covariance matrix is also independent of this choice, i.e.  of the selected function $\tilde{g}$. The additional constraints are only introduced to solve the identification issue. 

\medskip
{\bf Proposition 6} 

The asymptotic distribution of the IML estimator of matrix $A$ does not depend on the benchmark $\alpha_0^*$ in the identified set, i.e. on the additional optimization criterion.

\medskip
{\bf Proof:}

\noindent The asymptotic first-order conditions involve $J_0 \sqrt{T} (\hat{\theta}_T - \alpha^*_0)$ $+D_1 \sqrt{T} (\hat{\lambda}_T - \lambda^*_{10})+D_2 \sqrt{T} (\hat{\lambda}_{2T} - \lambda^*_{20})$ and $D'_1 \sqrt{T} (\hat{\alpha}_T - \alpha^*_0)$ $+D'_2 \sqrt{T} (\hat{\alpha}_T - \alpha^*_0)$ in the left hand side of system (4.13). 
Asymptotically, a change of the benchmark modifies the matrix $D_2$ as well as the associated Lagrange multipliers by linear transformations $R, R^{-1}$, respectively. Then, the first-order
condition provide the same solution for $\sqrt{T} (\hat{\alpha}_T - \alpha^*_0)$ when $D_2$ is replaced by $\tilde{D}_2 = D_0 R$ and $\hat{\lambda}_{2T} - \lambda^*_{20}$ by $\hat{\tilde{\lambda}}_{2T} - \tilde{\lambda}^*_{20} = R^{-1} (\hat{\lambda}_{2T} - \lambda^*_0)$, where $R$ is invertible. This proves that the asymptotic variance covariance matrix is independent of the choice of the additional optimization criterion.

\hfill Q.E.D.

\noindent We have introduced an IML estimator of $A_0$ under the non-negative rank restriction $Rk_+ (A_0) = K_0$, derived its asymptotic Gaussian behavior \footnote{This asymptotic Gaussian distribution is degenerate because of the reduced rank.} and the expression of the associated efficiency bound. 
The asymptotic variance-covariance matrix of $\hat{A}_T$ is obtained by applying the Slutsky formula based on the first-order expansions of $a \sum_{k=1}^K \pi_k \beta_k^* \gamma_k^*$ in a neighbourhood of $\alpha_0^* = (a_0^*, \pi_{0k}^*, \beta_{0k}^*, \gamma_{0k}^*, \; k=1,...,K)$. 
Then, the asymptotic confidence intervals that are identifiable on identifiable parameter functions of $A$ also have the asymptotic optimality properties.

%

\section{Extension to Nonparametric Identification}
\setcounter{equation}{0}\def\theequation{5.\arabic{equation}}

\subsection{The Identified Set}

The proof of Proposition 1 is valid in a nonparametric framework. Let us consider a pair of real variables $(X,Y)$ with a continuous joint distribution on a product of intervals and assume a positive pdf $f(x;y)$ on that interval. We can be interested in a decomposition~:

\begin{equation}
 f(x,y) = \Sum^K_{k=1} \pi_k \beta_k (x) \gamma_k (y), \tag{$1$}
\end{equation}

\noindent where $(\pi_k)$ defines the latent heterogeneity distribution, $\beta_k (x), \gamma_k (y)$ some pdf's for $x$ and $y$, respectively. This is a mixture model in which each joint distribution in the mixture satisfies the independence condition [see Compiani, Kitamura (2016) for mixtures in Econometrics] \footnote{ This condition can be written for more than two variables, so that the mixture becomes identifiable and  easier 
to analyse [Hall, Zhou (2003), Kasahara, Shimotsu (2009)]. A suitable
notion of rank has not yet been introduced for 3- or 4-entry tables.}.

The proof remains valid if functions $\beta_1, \ldots, \beta_K$ (resp. $\gamma_1, \ldots, \gamma_K)$ are a.s. linearly independent. Since the model is already normalized, Proposition 1 becomes~:\vspace{1em}

\textbf{Proposition 1'~:} Let us consider a true factorization of the joint pdf~:

$$
\begin{array}{lcl}
f_0 (x,y) & = & \Sum^K_{k=1} b_{0k} (x) \gamma_{0k} (y) =B_0 (x) C_0 (y),
\end{array}
$$

\noindent where the terms $b_{0k}$ are positive densities (not necessarily with unit mass) and $\gamma_{0k}$ are probability densities. Then, the other observationally equivalent decompositions are $B(x) = B_0 (x) Q, C(y) = C_0 (y) (Q')^{-1},$ where the matrix $Q$ is invertible, with unitary diagonal elements and such that~:

$$
B_0 (x) Q \geq 0, \; \mbox{a.s.}, \;\; C_0 (y) (Q')^{-1} \geq 0\; \mbox{a.s.}
$$

\vspace{1em}
\noindent The decomposition (1) of the joint pdf has alternative interpretations. For instance, the conditional p.d.f. can be written as:

$$
f(y|x) = \Frac{f(x,y)}{f(x)} = \Sum^K_{k=1} \Frac{b_{0k} (x)}{f(x)} \gamma_{0k} (y) \equiv \Sum^K_{k=1} b^*_{0k} (x) \gamma_{0k} (y).
$$

\noindent Therefore, it is equivalent to impose a reduced rank condition on the joint distribution, or on the conditional distribution. This problem is considered in Henry , Kitamura, Salanie (2014), where the pair of variables is denoted by $Y, W$,  the representation is written conditionally on a third variable $X$, and the independence condition on the elements of the mixtures is called the exclusion restriction [see also Compiani, Kitamura (2016)].

In general, the special case $K=2$ has to be solved directly [see e.g. Hall, Zhou (2003), Section 4].

\subsection{Nonparametric Identifying Maximum Likelhood}

The estimation approach outlined in Section 4 can be extended to the functional parameter framework, with functional parameters $b_k, k=1,\ldots, K$, scalar parameters $\pi_k, k=1,\ldots, K$, and additional parameters in $vec^* Q$.

Let us assume i.i.d. observations on $(X_i, Y_i), i=1,\ldots, n$. Then, a kernel based AML can be applied with the objective function~:

$$
\Sum^n_{i=1} \{ K (\Frac{x_i - x}{b_n}) K (\Frac{y_i - y}{b_n}) \log \Sum^K_{k=1} [b_k (x) \gamma_k (y)]\},
$$

\noindent where $K (.)$ denotes a kernel, $b_n$ the bandwidth and $\int \gamma_k (y) dy =1, \; k=1,...,K$. The maximisation with respect to $b_k (x), \gamma_k (y), k=1, \ldots, K$ provides the functional estimators of $b_k (.), \gamma_k (.),$ $k=1,\ldots, K$, and then functional estimators of $\pi_k, \beta_k (.), \gamma_k (.)$.\vspace{1em}

Next, due to the identification issue, an additional optimisation has to be performed to fix a mixture in the identified set. Criteria equivalent to the criteria introduced in Section 4.3.2 can be used, either the concentration criterion $\Sum^K_{k=1} (\pi_k \log \pi_k)$, or the alternative criterion $\det \Gamma$, with

$$
\Gamma = \Int \gamma (y) \gamma' (y) dy.
$$

The analysis of the asymptotic properties of this functional estimator approach are out of the scope of this paper and left for future research.

\section{Concluding  Remarks}

Although the nonnegative matrix factorization (NMF) is a well-known technique of dimension reduction for nonnegative matrices, it is used in the absence of an associated probability model for the observed data. Our paper fills this gap by considering a structural dynamic network model. We suggest new estimation methods for the set of identifiable NMF's and derive the asymptotic distribution of the estimated set and of specific elements of that set. Moreover, we provide a ML estimator of the non-negative matrix $A_0$ under a given non-negative rank constraint. The proposed approach is related to the nonparametric identification in mixture models.

Our approach can be used in a variety of applications with a lack of local identifiability. When an element of the identified set is characterized as a solution of an auxiliary optimization, the identified set can be parametrized given this element considered as a new origin. In practice, the dimension of the identified set can be very large and cannot be represented in a low-dimensional figure. However, it is possible to illustrate various elements or cuts of that set, which have structural interpretations and are easier to represent and discuss.

A typical partial identification of the same type is encountered in the analysis of the effects of monetary 
policy shocks under sign restrictions on the impulse response functions [Uhlig (2005)]. For example, in  a VAR(1) model $Y_t = \Phi Y_{t-1}+ u_t$, with $u_t \sim N(0, \Sigma)$ and $\Sigma = AA'$. The  matrix $A$ is not identifiable. The identified set can be reduced by imposing a restriction so that a "monetary policy impulse vector (a column of $A$) implies negative responses on prices and nonborrowed reserves and positive responses on federal funds rate, at all horizons" [Assumption A.1, Uhlig (2005)]. Then, there is a reduced identified set of these impulse vectors that can be easily parametrized. The idea of determining first the specific elements of the identified set by means of an additional optimization, before performing an extreme bounds analysis in the spirit of Leamer (1983) appears in Uhlig (2005), page 388 and Appendix B.2.

The standard factor analysis of time series by SVD is commonly followed by an interpretation of the dynamic factors. The factorial directions are projected on some observable time series to provide economic or financial interpretations of the dynamic factors (the so-called mimicking factors). A similar approach can be applied in our framework with partial identification. This may lead to selecting the most interpretable $\gamma_k's$.


\newpage
\centerline{\textbf{Online Appendix 1}}\vspace{1em}

\centerline{\textbf{Comparison of the Decompositions}}\vspace{1em}

\textbf{1. The NMF}\vspace{1em}

After the transformation the new NMF is~:

$$
\begin{array}{llcl}
 & A &=& \tilde{\beta}_1 \tilde{\gamma}'_1 + \tilde{\beta}_2 \tilde{\gamma}'_2, \\ \\
 \mbox{with} & \tilde{\beta}_1 &=& \beta_1 + q_{21} \beta_2 = q_{12} \beta_1 + \beta_2, \\ \\
 &\tilde{\gamma}_1 & =& \Frac{1}{1-q_{12} q_{21}} (\gamma_1 - q_{12} \gamma_2), \tilde{\gamma}_2 = \Frac{1}{1-q_{12} q_{21}} (-q_{21} \gamma_1 + \gamma_2).
\end{array}
$$

It is easily checked that $A = \beta_1 \gamma'_1 + \beta_2 \gamma'_2$.\vspace{1em}

\textbf{2. The decomposition (2.13).}\vspace{1em}

To get the new decomposition (2.13)~:

$$
A = a (\tilde{\pi} \tilde{\beta}^*_1 \tilde{\gamma}^{*'}_1 + (1-\tilde{\pi}) \tilde{\beta}^*_2 \tilde{\gamma}^{*'}_2],
$$

\noindent the new factorial directions have to be normalized with components summing up to one. We get~:

$$
\begin{array}{rcl}
\tilde{\beta}^*_1 & = & (\beta_1 + q_{21} \beta_2) / (\beta'_1 e + q_{21} \beta'_2 e), \\ \\
\tilde{\beta}^*_2 & = & (q_{12} \beta_1 + \beta_2) / (q_{12} \beta'_1 e +  \beta'_2 e), \\ \\
\tilde{\gamma}^*_1 & = & (\gamma_1 - q_{12} \gamma_2) / (\gamma'_1 e - q_{12} \gamma'_2 e), \\ \\
\tilde{\gamma}^*_2 & = & (-q_{21} \gamma_1 + \gamma_2) / (-q_{21} \gamma'_1 e + \gamma'_2 e), \\ \\
\tilde{\pi}/(1-\tilde{\pi}) & = & (\beta'_1 e + q_{21} \beta'_2 e) (\gamma'_1 e -q_{12} \gamma'_2 e)/ [(q_{12} \beta'_1 e + \beta'_2 e) [-q_{21} \gamma'_1 e + \gamma'_2 e)].
\end{array}
$$

\newpage
\centerline{\textbf{Online Appendix 2}}\vspace{1em}

\centerline{\textbf{Additional Assumptions for Asymptotic Results}}\vspace{1em}

\textbf{1. Consistency}\vspace{1em}

We provide below a set of additional assumptions a.1 to get the consistency. They require some uniform convergence of the objective function on the set $\mathcal{A}^*$ of all possible $\alpha$ on which the optimization is performed.\vspace{1em}

\textbf{Assumption a.1~:}\vspace{1em}

i) The set $\mathcal{A}^*$ is compact.

ii) $\log l(y_t | y_{t-1}; \alpha)$ is integrable for all $\alpha \in \mathcal{A}^*$.

iii) $\mathcal{A}_0 \subset A^*$.

iv) Uniform convergence of the objective function~:

$$
\begin{array}{l}
\sup_{\alpha \in \mathcal{A}^*} |\Frac{1}{T} \Sum^T_{t=1} \log l(y_t |y_{t-1}; \alpha) - E_0 \log l (y_t |y_{t-1};\alpha)|\\ \\
= 0_P (1/\sqrt{T}),
\end{array}
$$

\noindent where $E_0$ is the expectation with respect to the stationary distribution of $(Y_{t-1}, Y_t)$.

v) $\lim_{p\rightarrow \infty} \Frac{1}{T} \Sum^T_{t=1} \log l (y_t |y_{t-1}; \hat{\alpha}_T (\alpha_0, p)) = \max_{\alpha \in \mathcal{A}^*} \Frac{1}{T} \Sum^T_{t=1} \log l (y_t|y_{t-1};\alpha).$\vspace{1em}

The three first conditions are standard and used to prove the convergence of the set of solutions of the finite sample optimisations to $\mathcal{A}_0$. They imply condition $C_1$ on Chernozhukov et al. (2007) for instance (See also this reference for a proof). This is the convergence result in Proposition 3. Proposition 4 follows since this convergence of sets is uniform.

The last condition iv) is usually not introduced. It concerns the algorithm used to approximate solutions of the finite sample optimisation problems. This condition explains why we have introduced stronger conditions as in Assumption A.4 on the concavity of the log-likelihood function.

By construction the domain for $\pi, \beta^*_k, \gamma^*_k, k=1,\ldots, K$ is compact and its bounds as $\pi_k = 0,$ for some $k$, or $\beta_k = 0$, for some $k$ cannot be reached due to the rank condition. Therefore assumption a.1 i) concerns mainly scalar parameter $a$. \vspace{1em}

\textbf{2. Asymptotic Normality}\vspace{1em}

When $T$ tends to infinity, the estimator $(\hat{\alpha}_T, \hat{q}_T = q (\hat{\alpha}_T))$ will tend to $(\alpha^*_0, 0)$. Let us assume~:\vspace{1em}

\textbf{Assumption a.2~:} \vspace{1em}

i) $\alpha^*_0$ is in the interior of the true set $\mathcal{A}_0$.

ii) $0$ is in the interior of the set $Q (\alpha^*_0)$ of admissible values of $q$  constructed from $\alpha^*_0$.

iii) The log-likelihood function is twice continuously differentiable with respect to $\alpha$.

iv) The additional objective function $\tilde{g}$ is continuously differentiable with respect to $q$ and continuously cross differentiable with respect to $q$ and $\alpha$.

v) The function $q(.)$ exists and is continuously differentiable

vi) The score $\Frac{\partial \log l}{\partial \alpha}  (y_t| y_{t-1}; \alpha)$ has second-order moments.

vii) The Hessian $\Frac{\partial^2 \log l}{\partial \alpha \partial \alpha'} (y_t | y_{t-1}; \alpha)$ has first-order moments.

viii) The matrix $ (Id-P) J_0$ has rank $\dim \alpha - \dim q - 1-2K$.\vspace{1em}

\end{document}